\documentclass[12pt]{article}
\usepackage{times,amsfonts,amssymb,amsmath,bbm,stmaryrd,amscd,latexsym}
\newcommand{\cw}{}

\newcommand{\cg}{}
\newcommand{\cy}{}

\newcommand{\ccy}{}
\def\a{\alpha}

\def\th{\theta}

\def\Om{\Omega}

\def\ve{\varepsilon}

\newcommand{\De}{{\Delta\vphantom{\big|}}}
\renewcommand{\gg}{g}
\newcommand{\el}{\ell}
\def\sfrac#1#2{{\textstyle\frac{#1}{#2}}}

\def\half{{\sfrac12}}

\def\ssum#1{{\textstyle\sum_{#1}}}

\def\im{{\mathrm i}}

\def\ep{{\mathrm e}}

\def\pa{\partial}
\def\>{\rangle}
\def\<{\langle}
\def\+{{\dagger}}
\def\={\ =\ }

\newcommand{\E}{{\scriptscriptstyle E}}
\newcommand{\wb}{{\bar{w}}}
\newcommand{\rb}{{\bar{\rho}}}
\newcommand{\cD}{{\cal D}}
\newcommand{\cH}{{\cal H}}
\newcommand{\cL}{{\cal L}}
\newcommand{\cM}{{\cal M}}

\newcommand{\res}{{\mathrm{res}}}

\newcommand{\R}{\mathbb R}
\newcommand{\Z}{\mathbb Z}
\newcommand{\NN}{{\mathbb{N}}}

\newcommand{\beq}{\begin{equation}}
\newcommand{\eeq}{\end{equation}}
\newcommand{\bea}{\begin{eqnarray}}
\newcommand{\eea}{\end{eqnarray}}
\newcommand{\bal}{\begin{aligned}}
\newcommand{\eal}{\end{aligned}}

\newcommand{\with}{{\quad\textrm{with}\quad}}
\newcommand{\for}{{\quad\textrm{for}\quad}}
\renewcommand{\and}{{\quad\textrm{and}\quad}}
\newcommand{\und}{\qquad\textrm{and}\qquad}

\newcommand{\comma}{\quad{,}\qquad}
\begin{document}
\begin{titlepage}
\begin{flushright}      
ITP--UH--08/16\\        
\end{flushright}        

\vskip 2.0cm

\begin{center}
{\Large\bf The tetrahexahedric Calogero model}

\vskip 1.5cm

{\Large 
Francisco Correa {\large \ \ and \ \ } Olaf Lechtenfeld
}

\vspace{0.5cm}

\noindent 
{\it
Institut f\"ur Theoretische Physik and Riemann Center for Geometry and Physics, \\
Leibniz Universit\"at Hannover,
Appelstrasse 2, 30167 Hannover, Germany}



\vspace{2.0cm}

\begin{abstract}
\noindent
We consider the spherical reduction of the rational Calogero model
(of type $A_{n-1}$, without the center of mass)
as a maximally superintegrable quantum system.
It describes a particle on the $(n{-}2)$-sphere in a very special potential. 
A detailed analysis is provided of the simplest non-separable case, $n{=}4$, 
whose potential blows up at the edges of a spherical tetrahexahedron,
tesselating the two-sphere into 24 identical right isosceles spherical
triangles in which the particle is trapped.
We construct a complete set of independent conserved charges and of 
Hamiltonian intertwiners and elucidate their algebra.
The key structure is the ring of polynomials in Dunkl-deformed
angular momenta, in particular the subspaces invariant
and antiinvariant under all Weyl reflections, respectively.
\end{abstract}

\end{center}

\vfill

Talk presented at SQS-15 during 03-08 August, 2015, at JINR, Dubna, Russia

\end{titlepage}


\newpage

\section{Some history}

The Calogero model has a 45-year history, starting in 1971 with the original Calogero paper~\cite{cal71}.
Ten years later Olshanetsky and Perelemov generalized the $A_{n-1}$ model to arbitrary finite-dimen\-sional
Lie algebras and demonstrated their classical~\cite{olpe81} and quantum~\cite{olpe83} integrability.
In 1983, the superintegrability of the Calogero-Moser system was established by Wojciechowski~\cite{woj83}.
Starting with their seminal 1990 paper~\cite{vech90} on commutative rings of partial differential operators 
and Lie algebras, Veselov and Chalykh initiated a series of works on intertwiners (shift operators) and 
the exact energy spectrum for integer couplings (multiplicities). In parallel, employing the 
differential-difference operators associated to reflection groups and introduced by Dunkl~\cite{dun89},
Heckman gave an elementary construction for commuting charges and intertwiners~\cite{hec91}.
The first investigation of the spherical reduction of the rational Calogero model (here called
`angular Calogero model') goes back to M.~Feigin in 2003~\cite{fei03}. The $A_2$ and $A_3$~cases were
analyzed classically in 2008 by Hakobyan, Nersessian and Yeghikyan~\cite{haneye08}, and five years later
the quantum energy spectra and eigenstates were derived for all angular Calogero models by M.~Feigin,
Lechtenfeld and Polychronakos~\cite{felepo13}. More recently, M.~Feigin and Hakobyan presented a deeper
analysis of the algebra of Dunkl angular momentum operators, and just now the $A_2$ and $A_3$ angular
models have been reconsidered on the quantum level by the authors~\cite{cole15}. 
This talk reviews their results.

\section{The angular (relative) Calogero model}

In the first half of the talk, let us introduce the spherical reduction of
rational $A_{n-1}$ Calogero model and present some of its salient features.
In an $n$-particle quantum phase space with particle coordinates $x^\mu$ and
momenta $p_\mu$, where $\mu=1,2,\ldots,n$, subject to 
$[x^\mu,p_\nu] = \im\,\delta^\mu_{\ \nu}$, the rational Calogero Hamiltonian
(after separating the center of mass) reads
\beq
H \= \sum_{\mu<\nu}^n \Bigl\{ \sfrac{1}{2n}(p_\mu{-}p_\nu)^2\ +\
\frac{\cg g(g{-}1)}{(x^\mu{-}x^\nu)^2} \Bigr\} \ .
\eeq
The strength of the inverse-square two-body potential is parametrized by a
real coupling constant $g$ (which could be taken $\ge\sfrac14$).
In the `relative' $2(n{-}1)$-dimensional phase space, a radial coordinate and momentum
are defined via
\beq 
\sfrac1n\sum_{\mu<\nu} (x^\mu{-}x^\nu)^2 \= r^2 \und
\sfrac1n\sum_{\mu<\nu} (p_\mu{-}p_\nu)^2 \= 
p_r^2 + \sfrac{1}{r^2}L^2 + \sfrac{(n-2)(n-4)}{4\,r^2}\ .
\eeq
It is convenient to switch to $n{-}1$ `relative' coordinates $y^i$ and momenta $p_i$,
with $i=1,2,\ldots,n{-}1$,
\beq 
r^2 \= \sum_{i=1}^{n-1} (y^i)^2 \quad\cw,\cy\quad p_i\equiv p_{y^i} \quad\cw,\cy\quad
L_{ij} \= -\im(y^i p_j-y^j p_i) \quad\cw,\cy\quad L^2 \= -\sum_{i<j}L_{ij}^2\ .
\eeq
In terms of polar coordinates $(r,\vec\th)$ on $\R^{n-1}$, 
the Hamiltonian takes the form
\beq 
H \= \sfrac12 p_r^2\ +\ \sfrac{(n-2)(n-4)}{8\,r^2}\ +\ \sfrac{1}{r^2} H_\Om 
\qquad\textrm{with}\qquad
H_\Om \= \sfrac12 L^2  + U(\vec\th) \ ,
\eeq
where the angular potential is
\beq
U(\vec\th) \= r^2 \sum_{\mu<\nu} \frac{\cg g(g{-}1)}{(x^\mu{-}x^\nu)^2}
\ \=\ r^2 \sum_{\alpha\in{\cal R}_+} \frac{\cg g(g{-}1)}{(\alpha\cdot y)^2}
\ \=\ \sfrac{\cg g(g{-}1)}{2} \sum_{\alpha\in{\cal R}_+} \cos^{-2}\th_\alpha\ .
\eeq
Here, we introduced the $A_{n-1}$ positive root system ${\cal R}_+$ and 
the angle $\th_\alpha$ between the point $\vec\theta\in S^{n-2}$ and the root~$\alpha$.
$H_\Om$ is the angular (relative) Calogero Hamiltonian, our object of interest.

In the position representation, we pass to differential operators,
\beq
p_i\ \mapsto\ -\im\pa_i \qquad\cw\Longrightarrow\cy\qquad
p_r\ \mapsto\ -\im\bigl(\pa_r + \sfrac{n-2}{2\,r}\bigr)\ ,
\eeq
so our Hamilton operators become
\beq 
\begin{aligned}
H\ &\mapsto\ -\sfrac12 \bigl(\pa_r^2 + \sfrac{n-2}{r}\pa_r\bigr)\ +\ \sfrac{1}{r^2} H_\Om
\= S^{-1} \Bigl[ -\sfrac12 \bigl(\pa_r^2 - \sfrac{(n-2)(n-4)}{4\,r^2}\bigr)\ +\
\sfrac{1}{r^2} H_\Om \Bigr] \,S \\[8pt]
H_\Om &\mapsto\ -\half\sum_{i<j}\bigl( y^i\pa_j{-}y^j\pa_i\bigr)^2 \ +\
r^2 \sum_{\alpha\in{\cal R}_+} \frac{\cg g(g{-}1)}{(\alpha\cdot y)^2}
\qquad\qquad\qquad\cw\textrm{with}\quad S=r^{\frac{n-2}{2}}\ .
\end{aligned}
\eeq
The spectrum and the eigenfunctions of $H$ are known,
\beq 
\begin{aligned}
H\,\Psi_{\E,q} &\= E\,\Psi_{\E,q} \qquad\textrm{\cw with}\qquad E\in\R_{\ge0} \und \\[8pt]
\Psi_{\E,q}(r,\vec\th) &\= r^{-\frac{n-3}{2}}\,J_{q+(n-3)/2}({\scriptstyle\sqrt{2E}}\,r)\,v_q(\vec\th)\ ,
\end{aligned}
\eeq
where we took advantage of the conformal invariance to separate in polar coordinates. 
The angular wave function $v_q(\vec\th)$ is an eigenfunction of the angular Hamiltonian,
whose spectrum is also in the literature,
\beq
\begin{aligned}
H_\Om\,v_q &\= \ve_q\,v_q \quad\with\quad \ve_q \= \half\,q\,(q+n-3) \und \\[8pt]
q &\= \half n(n{-}1)\,\gg +\el \qquad\textrm{\cw where}\qquad
\el\=3\el_{\ccy 3}+4\el_{\ccy 4}+\ldots+n\el_{\ccy n} \ \in\NN_0\ .
\end{aligned}
\eeq
The degeneracy of energy level $\ve_q$ is given by
\beq
\textrm{deg}_n(\ve_q) \= p_n(\el)-p_n({\ccy\el{-}1})-p_n({\ccy \el{-}2})+p_n({\ccy \el{-}3})
\eeq
with the restricted partitions $p_n(\el)$ given by the simple generating function
\beq
p_n(t)\ :=\ \sum_{\el=0}^\infty p_n(\el)\,t^\el \= \prod_{m=1}^n \bigl(1-t^m\bigr)^{-1}\ .
\eeq
Relevant for this talk are the cases of $n{=}3$ and $4$,
\beq 
\begin{aligned}
\textrm{deg}_3(\el) &\= \begin{cases}
0 &\for \el=1,2 \ \textrm{mod} \ 3 \\
1 &\for \el=0 \ \textrm{mod}\ 3
\end{cases} \ ,\\
\textrm{deg}_4(\el) &\= \Bigl\lfloor\frac{\el}{12}\Bigr\rfloor \ +\ \begin{cases}
0 &\for \el=1,2,5\ \textrm{mod} \ 12 \\
1 &\for \el=\textrm{else} \ \textrm{mod} \ 12
\end{cases} \ .
\end{aligned}
\eeq
All the interesting nontrivial structure is hidden in the angular eigenfunctions:
\beq 
v_q(\vec\th) \ \equiv\ v_\el^{(\gg)}(\vec\th) \ \sim\
r^{n-3+q} \Bigl(\prod_{\mu=3}^n \sigma_\mu\bigl(\{\cD_i\}\bigr)^{\el_\mu}\Bigr)\,\De^\gg\,r^{3-n-n(n-1)\gg}\ ,
\eeq
which employs the Vandermonde~$\De$ and the (mutually commuting) Dunkl operators~$\cD_i$ 
as arguments in the $\mu$th Newton sum~$\sigma_\mu(y)=\sum_i (y^i)^\mu$,
\beq 
\De \= \prod_{\alpha\in{\cal R}_+} \alpha\cdot y \und
\cD_i \= \pa_i \ -\ \gg\sum_{\alpha\in{\cal R}_+} \frac{\alpha_i}{(\alpha\cdot y)^2}\,s_\alpha\ ,
\eeq
where $s_\alpha$ denotes the reflection on the hyperplane orthogonal to the root~$\alpha$.
These wave functions contain a factor of $\De^g$ and are directly related to 
Dunkl-deformed Weyl-symmetric harmonic polynomials,
\beq 
v_\el^{(\gg)}(\vec\th) \= r^{-{\cg q}}\,\De^\gg\,\widetilde{h}_\el^{(\gg)}
\quad\with\quad H\,\bigl(\De^\gg\,\widetilde{h}_\el^{(\gg)}\bigr) \= 0\ .
\eeq

The $\cD_i$, $y^i$ and $s_\alpha$ form a rational Cherednik algebra.
The restriction `res' of its elements to Weyl-invariant functions yields important differential operators,
in particular our Hamiltonians. To make this explicit, we `Dunkl-deform' not only the linear momenta,
$\pa_i\Rightarrow\cD_i$ but also the angular momenta,
\beq 
L_{ij} \ \mapsto\ -(y^i\pa_j{-}y^j\pa_i) \qquad\cw\Longrightarrow\cy\qquad
\cL_{ij} \= -(y^i\cD_j - y^j\cD_i)\ ,
\eeq
and define the `pre-Hamiltonians'
\beq 
\cH \=-\half\sum_i \cD_i^2 \und
\cH_\Om \= -\sfrac12\sum_{i<j}\cL_{ij}^2 \ +\
\sfrac12\,\gg\ssum{\a}s_\a\,(\gg\ssum{\a}s_\a+n{-}3)\ ,
\eeq
whose Weyl-symmetric restriction produce 
\beq
H \= \res(\cH) \ \ \und\ \
H_\Om \= \res(\cH_\Om) \= \half\res\bigl(\cL^2\bigr)\ +\ \ve_{\cg q}({\ccy \el{=}0})\ .
\eeq
The Cherednik subalgebra generated by the $\cL_{ij}$ and the Weyl reflections 
is given by the relations
\beq
[ \cL_{ij},\cL_{k\ell} ] \=
\cL_{i\ell}{\cal S}_{jk} - \cL_{ik}{\cal S}_{j\ell} - \cL_{j\ell}{\cal S}_{ik} + \cL_{jk}{\cal S}_{i\ell}\ ,
\eeq
\beq
\cL_{ij}\cL_{k\ell}+\cL_{jk}\cL_{i\ell}+\cL_{ki}\cL_{j\ell} \= 
\cL_{ij}{\cal S}_{k\ell}+\cL_{jk}{\cal S}_{i\ell}+\cL_{ki}{\cal S}_{j\ell}\ ,
\eeq
\beq
[{\cal S}_{ij},\cL_{k\ell}]=0 \comma
\{{\cal S}_{ij},\cL_{ij}\}=0 \comma
{\cal S}_{ij}\cL_{ik} \= \cL_{jk}{\cal S}_{ij}\ ,
\eeq
\beq
\textrm{with}\qquad
{\cal S}_{ij} \= \begin{cases}
-\gg\,s_{ij} & \for i\neq j \\
1+\gg\ssum{k(\neq i)} s_{ik} & \for i=j \end{cases}\ .
\eeq
It is a `Dunkl deformation' of $so(n{-}1)$, with $H_\Om$ being the Casimir invariant

A hallmark of Calogero models is their isospectrality, which is characterized by the existence
of intertwining (or shift) operators relating the energy spectra at couplings $g$ and $g{+}1$.
This concept is well established for the full rational model, but is also works in the angular
submodel. There, angular intertwiners are differential operators $M_s$ in $\vec\theta$ of some order~$s$,
constructed with the following recipe,
\beq 
M_s \= \res(\cM_s) \quad\with\quad
\cM_s \= \textrm{ Weyl antiinvariant in $\{\cL_{ij}\}$ of degree $s$ }
\eeq
Since $[\cL_{ij},\cH]=0$ and $M_s$ has no $r$~dependence, it follows that
\beq
\begin{aligned}
{}[ \cM_s,\cH_\Om ] = 0
\qquad\cw\Longrightarrow\cy\qquad
& M_s^{(\gg)} H_\Om^{(\gg)} \= H_\Om^{({\cg g+1})} M_s^{(\gg)} \\[4pt]
\und & M_s^{(\gg)} v_{\el}^{(\gg)}\ \sim\ v_{{\ccy \el-n(n-1)/2}}^{({\cg g+1})}
\end{aligned}
\eeq
The adjoint ${M_s^{(\gg)}}^\+ = M_s^{(-\gg)}$ intertwines in the opposite direction,
i.e.\ $M_s^{(-\gg)} v_{\el}^{({\cg g+1})} \sim v_{\el+n(n-1)/2}^{(\gg)}$.
It follows that for integer~$g$ we can obtain the angular eigenfunctions more directly by
successively applying intertwiners to the free eigenfunctions, say at $g{=}1$,
\beq
v_{\el}^{(\gg)}\ \sim\ M_{s_1}^{(\gg-1)}\,M_{s_2}^{(\gg-2)} \cdots 
M_{s_{g-1}}^{(1)}\,v_{\el+(g-1)n(n-1)/2}^{(1)}\ .
\eeq
An important issue is the existence of conserved charges beyond the Hamiltonian~$H_\Om$. 
Obviously, $[M_s^\+ M_s^{\phantom\+},H_\Om]=0=[M_s^{\phantom\+} M_s^\+,H_\Om]$, but
this need not provide new quantities. However, any Weyl-invariant polynomial ${\cal C}_t(\cL_{ij})$ 
of some degree~$t$ gives rise to a conserved charge,
\beq
{\cal C}_t(\cL_{ij}) \quad \textrm{Weyl-invariant}  \qquad\cw\Longrightarrow\cy\qquad
C_t = \res({\cal C}_t) \quad \textrm{commutes with} \ H_\Om\ .
\eeq
We already know of $C_0=1$ and $C_2=-\res\bigl(\cL^2\bigr)$ but expect $2n{-}5$
algebraically independent constants of motion (beyond $C_0$) in a superintegrable theory.
Other than the Liouville charges in the full Calogero model, they will generically mix
under the intertwining action,
\beq
M_s^{(g)}\,C_t^{(g)} \= \smash{\sum_{s',t'}} \Gamma_{st}^{s't'}(g)\,C_{t'}^{(g+1)} \,M_{s'}^{(g)}
\eeq
with some coefficient functions $\Gamma_{st}^{s't'}(g)$.

\section{Warmup: the hexagonal or P\"oschl-Teller model}

Let us illustrate the structures just mentioned on the first nontrivial example, 
which at $n{=}3$ is the $A_2$~model. Its spherical reduction (to the unit circle)
is known as the P\"oschl-Teller model, but we call it `hexagonal' because the 
potential is singular at angles $\phi=(2k{+}1)\pi/6$.
The relation between the 3 particle coordinates $x^\mu$
and the 2 Jacobi relative coordinates $y^i$ orthogonal to the center of mass~$X$ is
\beq
\begin{aligned}
x^1&\=X + \sfrac1{\sqrt{2}}\,y^1 + \sfrac1{\sqrt{6}}\,y^2  \comma &
\pa_{x^1}&\=\sfrac13\pa_X+ \sfrac1{\sqrt{2}}\,\pa_{y^1} + \sfrac1{\sqrt{6}}\,\pa_{y^2} \ ,\\
x^2&\=X - \sfrac1{\sqrt{2}}\,y^1 + \sfrac1{\sqrt{6}}\,y^2  \comma &
\pa_{x^2}&\=\sfrac13\pa_X- \sfrac1{\sqrt{2}}\,\pa_{y^1} + \sfrac1{\sqrt{6}}\,\pa_{y^2} \ ,\\
x^3&\=X - \sfrac2{\sqrt{6}}\,y^2 \comma & 
\pa_{x^3}&\=\sfrac13\pa_X  - \sfrac2{\sqrt{6}}\,\pa_{y^2} \ .
\end{aligned}
\eeq
Performing the polar decomposition and introducing a complex coordinate,
\beq 
y^1 \= r\,\cos\phi \and y^2 \= r\,\sin\phi
\qquad\cw\Longrightarrow\cy\qquad w\ :=\ y^1+\im y^2 \= r\,\ep^{\im\phi}\ ,
\eeq
the angular Hamiltonian takes the form
\beq 
H_\Om \= \half\bigl(w\pa_w-\wb\pa_\wb\bigr)^2\ +\ {\cg g(g{-}1)}\frac{18\,(w\wb)^3}{(w^3+\wb^3)^2}
\qquad\qquad\qquad\textrm{since}
\eeq
\beq 
U(\phi) \=
\sfrac{\cg g(g-1)}2 \sum_{k=0,1,2} \cos^{-2}(\phi{+}k\sfrac{2\pi}{3})
\= \sfrac92 {\cg g(g{-}1)}\,\cos^{-2}(3\phi)
\= {\cg g(g{-}1)}\sfrac{18\,(w\wb)^3}{(w^3+\wb^3)^2}\ .
\eeq
Its spectrum depends on a single quantum number $\ell=3\ell_3$, with $\ell_3\in\NN_0$,
\beq 
\ve_q \= \half q^2 \qquad\textrm{\cw with}\qquad
q \= 3\gg+\el \= 3(\gg+\el_{\ccy 3}) \und \textrm{deg}(\ve_q)=1\ .
\eeq
Since the third Newton sum is $\sigma_3(w,\wb)=w^3-\wb^3$,
the angular wave functions are constructed as
\beq 
v_q(\phi) \ \equiv\ v_\el^{(\gg)}(\phi) \ \sim\
r^{\cg q}\,\bigl(\cD_w^3-\cD_\wb^3\bigr)^{\el_{\ccy 3}}\,\De^\gg\,r^{-6\gg} \=
r^{-\cg q}\,\De^\gg\,\widetilde{h}_\el^{(\gg)}(w^3,\wb^3)\ ,
\eeq
where the ingredients are
\beq 
\De\ \sim\ w^3+\wb^3 \ \sim\ r^3\,\cos(3\phi)
\qquad\qquad\qquad\textrm{and}
\eeq
\beq 
\cD_w \= \pa_w\ -\ g\,\Bigl\{
\frac{1}{w+\wb}\,s_0 + \frac{\rho}{\rho w+\rb\wb}\,s_+ + \frac{\rb}{\rb w+\rho\wb}\,s_- \Bigr\}
\with \rho=\ep^{2\pi\im/3}\ .
\eeq
The application of the Dunkl operators can be evaluated analytically, arriving at
\beq 
\widetilde{h}_\el^{(\gg>0)}(w^3,\wb^3) \=
\sum_{k=0}^{\el_{\ccy 3}} (-1)^k\,\frac{\Gamma(\gg{+}k)\,\Gamma(\gg{+}\el_{\ccy 3}{-}k)}
{\Gamma(\gg)\,\Gamma(1{+}k)\,\Gamma(1{+}\el_{\ccy 3}{-}k)} \ w^{\el-3k}\wb^{3k}\ .
\eeq
The table below lists some low-lying hexagonal wave functions, abbreviating
$(m\,\bar{m}):=w^{3m}\wb^{3\bar{m}}$.
\begin{center}
{\small
\begin{tabular}{|c|ccc|}
\hline
$\el$ & $\widetilde{h}_\el^{(\cg 0)}$ & $\widetilde{h}_\el^{(\cg 1)}$ & $\widetilde{h}_\el^{(\cg 2)} \phantom{\Big|}$ \\ 
\hline
$\phantom{\Big|}0$ & 
$(00)$ & $(00)$ & $(00)$  \\
$\phantom{\Big|}3$ & 
$(10)-(01)$ & $(10)-(01)$  & $(10)-(01)$  \\
$\phantom{\Big|}6$ & 
$(20)+(02)$ & $(20)-(11)+(02)$ &  $3(20)-4(11)+3(02)$  \\
$\phantom{\Big|}9$ & 
$(30)-(03)$ & $(30)-(21)+(12)-(03)$ & $4(30)-6(21)+6(12)-4(03)$  \\
$\phantom{\Big|}12$ & 
$(40)+(04)$ & $(40)-(31)+(22)-(13)+(04)$ & $5(40)-8(31)+9(22)-8(13)+5(04)$  \\
\hline
\end{tabular}
}
\end{center}
The simplest Weyl antiinvariant build from $\cL_{12}$ is the Dunklized angular momentum itself,
\beq
\begin{aligned}
\cM_1\ &\sim\ \im \bigl( w\cD_w-\wb\cD_\wb \bigr) \\[4pt]
&\sim\ \im \bigl( w\pa_w-\wb\pa_\wb \bigr)\ -\ \im\,\gg\,\Bigl\{
\frac{w-\wb}{w+\wb}\,s_0 +\frac{\rho w-\rb\wb}{\rho w+\rb\wb}\,s_+
+\frac{\rb w-\rho\wb}{\rb w+\rho\wb}\,s_-\Bigr\}\ ,
\end{aligned}
\eeq
whose Weyl-symmetric restriction gives a most simple angular intertwiner,
\beq
M_1\ \sim\ \im \bigl( w\pa_w-\wb\pa_\wb \bigr)\ -\ 3\im\,\gg\,\frac{w^3-\wb^3}{w^3+\wb^3} \=
\im\,\Delta^{\gg}\bigl( w\pa_w-\wb\pa_\wb \bigr) \Delta^{-\gg} \=
\pa_\phi + 3\,\gg\,\tan3\phi\ ,
\eeq
which allows for an even simpler recursion relation for the hexagonal wave functions,
\beq
\widetilde{h}_\el^{(\gg+1)} \ \sim\ \im\,\Delta^{-1}\bigl( w\pa_w-\wb\pa_\wb \bigr)\ \widetilde{h}_{\el+3}^{(\gg)}\ .
\eeq
Iterating this recursion is an easier way to construct these wave functions from the ground state.

Because
\beq 
\bigl(M_1^\+ M_1^{\phantom\+}\bigr)^{(\gg)} \= -2\,H_\Om^{(\gg)} + 9\,\gg^2
\= -\res(\cL^2) \= -C_2^{(\gg)}\ ,
\eeq
there is no further conserved charge besides the angular Hamiltonian in the hexagonal model.

\section{Tetrahexahedric model: the spectrum}

Now we pass to the next and more interesting case, $n{=}4$. 
This angular model is quite new and describes a particle on the two-sphere with a non-separable potential.
We call it tetrahexahedric because the singular loci of the potential are six great circles which form the
edges of a spherical polyeder called tetrahexahedron. Therefore, the particle is trapped in one of 24
identical fundamental domains (the faces), which have the shape of a (spherical) right isosceles triangle.
It is convenient to pass to Walsh-Hadamard relative coordinates (due to $A_4\simeq D_3$):
\beq
\begin{aligned}
x^1\=X+\half(+x+y+z) \comma &
\pa_{x^1}\=\sfrac14\pa_X+\half(+\pa_{x}+\pa_{y}+\pa_{z})\ , \\
x^2\=X+\half(+x-y-z) \comma &
\pa_{x^2}\=\sfrac14\pa_X+\half(+\pa_{x}-\pa_{y}-\pa_{z})\ , \\
x^3\=X+\half(-x+y-z) \comma &
\pa_{x^3}\=\sfrac14\pa_X+\half(-\pa_{x}+\pa_{y}-\pa_{z})\ , \\
x^4\=X+\half(-x-y+z) \comma &
\pa_{x^4}\=\sfrac14\pa_X+\half(-\pa_{x}-\pa_{y}+\pa_{z})\ ,
\end{aligned}
\eeq
and introduce spherical coordinates
\beq
x \= r\,\sin\th\,\cos\phi \comma
y \= r\,\sin\th\,\sin\phi \comma
z \= r\,\cos\th\ .
\eeq
The angular momenta and the spherical Laplacian take the familiar form
\beq 
L_x = -(y\pa_z{-}z\pa_y) \ \cw,\cy\qquad
L_y = -(z\pa_x{-}x\pa_z) \ \cw,\cy\qquad
L_z = -(x\pa_y{-}y\pa_x)
\eeq
\beq
\und 
L^2 \= -(L_x^2+L_y^2+L_z^2) \=
-\sfrac{1}{\sin\th}\pa_\th\sin\th\,\pa_\th - \sfrac{1}{\sin^2\th}\pa_\phi^2\ ,
\eeq
and the angular Hamiltonian reads \ $H_\Om=\half L^2+U(\th,\phi)$ \ with
\bea 
U(\th,\phi) \!\!&=&\!\! 2\,{\cg g(g{-}1)}\bigl(x^2+y^2+z^2) \Bigl(
\frac{x^2+y^2}{(x^2-y^2)^2} + \frac{y^2+z^2}{(y^2-z^2)^2} + \frac{z^2+x^2}{(z^2-x^2)^2}\Bigr) \\[4pt]
&=&\!\! 2{\cg g(g{-}1)}\biggl\{
\frac1{\sin^2\th\,\cos^22\phi}+
\frac{\cos^2\th+\sin^2\th\cos^2\phi}{(\cos^2\th-\sin^2\th\cos^2\phi)^2}+
\frac{\cos^2\th+\sin^2\th\sin^2\phi}{(\cos^2\th-\sin^2\th\sin^2\phi)^2} \biggr\} \notag
\eea
The tetrahexahedric energy spectrum is given by
\beq
\ve_q \= \half q\,(q{+}1) \qquad\textrm{\cw with}\qquad q \= 6\gg+\el \= 6\gg+3\el_{\ccy 3}{+}4\el_{\ccy 4}
\und \ell_3,\ell_4\in\NN_0\ .
\eeq
The corresponding wave functions can be computed from
\beq
v_\el^{(\gg)}(\th,\phi)\ \sim\
r^{{\cg q}+1}\,\bigl(\cD_x\cD_y\cD_z\bigr)^{\el_{\ccy 3}}\,
\bigl(\cD_x^4{+}\cD_y^4{+}\cD_z^4\bigr)^{\el_{\ccy 4}}\,\De^\gg\,r^{1-12\gg}
\= r^{-{\cg q}}\,\De^\gg\,\widetilde{h}_\el^{(\gg)}(x,y,z)
\eeq
\beq
\textrm{with}\qquad \De \= (x^2-y^2)(x^2-z^2)(y^2-z^2)
\eeq
and the linear Dunkl operators
\beq 
\begin{aligned}
\cD_x&\=\pa_x\ -\ \sfrac{\gg}{x{+}y}\,s_{x+y}-\sfrac{\gg}{x{-}y}\,s_{x-y}-
\sfrac{\gg}{z{+}x}\,s_{x+z}-\sfrac{\gg}{x{-}z}\,s_{z-x} \ ,\\
\cD_y&\=\pa_y\ -\ \sfrac{\gg}{y{+}x}\,s_{x+y}-\sfrac{\gg}{y{-}x}\,s_{x-y}-
\sfrac{\gg}{y{+}z}\,s_{y+z}-\sfrac{\gg}{y{-}z}\,s_{y-z} \ ,\\
\cD_z&\=\pa_z\ -\ \sfrac{\gg}{z{+}x}\,s_{z+x}-\sfrac{\gg}{z{-}x}\,s_{z-x}-
\sfrac{\gg}{z{+}y}\,s_{y+z}-\sfrac{\gg}{z{-}y}\,s_{y-z}
\end{aligned}
\eeq
including the elementary reflections constituting the $S_4$ Weyl group action,
\beq 
\begin{aligned}
& s_{x+y} : \ (x,y,z)\mapsto(-y,-x,+z) \comma
  s_{x-y} : \ (x,y,z)\mapsto(+y,+x,+z) \ ,\\
& s_{y+z} : \ (x,y,z)\mapsto(+x,-z,-y) \comma
  s_{y-z} : \ (x,y,z)\mapsto(+x,+z,+y) \ ,\\
& s_{z+x} : \ (x,y,z)\mapsto(-z,+y,-x) \comma
  s_{z-x} : \ (x,y,z)\mapsto(+z,+y,+x) \ .
\end{aligned} 
\eeq
The following table lists the low-lying tetrahexahedric wave functions for $g{=}0$ and $g{=}1$, 
using the notation \ $\{rst\}:=x^ry^sz^t+x^ry^tz^s+x^sy^tz^r+x^sy^rz^t+x^ty^rz^s+x^ty^sz^r$.
\begin{center}
{\small
\begin{tabular}{|cc|c|}
\hline
$\el_{\ccy 3}$ & $\el_{\ccy 4}$ & $\widetilde{h}_{\el_{\ccy 3},\el_{\ccy 4}}^{({\cg 0})}\phantom{\Big|}$ \\ 
\hline
0 & 0 & $\{000\}$ \\
1 & 0 & $\{111\}$ \\
0 & 1 & $\{400\}-3\{220\}$ \\
2 & 0 & $\{600\}-15\{420\}+30\{222\}$ \\
1 & 1 & $3\{511\}-5\{331\}$ \\
0 & 2 & $\{800\}-28\{620\}+35\{440\}$ \\
3 & 0 & $9\{711\}-63\{531\}+70\{333\}$ \\
2 & 1 & $\{1000\}-45\{820\}+42\{640\}+504\{622\}-630\{442\}$ \\
1 & 2 & $5\{911\}-60\{731\}+63\{551\}$ \\
4 & 0 & $36\{1200\}-2376\{1020\}+2445\{840\}+46125\{822\}+4893\{660\}-215250\{642\}+179375\{444\}$ \\
0 & 3 & $101\{1200\}-6666\{1020\}+47100\{840\}+8685\{822\}-42609\{660\}-40530\{642\}+33775\{444\}$ \\
\hline
\end{tabular}
}
\end{center}
\begin{center}
{\small
\begin{tabular}{|cc|c|}
\hline
$\el_{\ccy 3}$ & $\el_{\ccy 4}$ & $\widetilde{h}_{\el_{\ccy 3},\el_{\ccy 4}}^{({\cg 1})}\phantom{\Big|}$ \\
\hline
0 & 0 & $\{000\}$ \\
1 & 0 & $\{111\}$ \\
0 & 1 & $3\{400\}-11\{220\}$ \\
2 & 0 & $3\{600\}-39\{420\}+196\{222\}$ \\
1 & 1 & $5\{511\}-13\{331\}$   \\
0 & 2 & $\{800\}-20\{620\}+23\{440\}+12\{422\}$ \\
3 & 0 & $3\{711\}-27\{531\}+56\{333\}$  \\
2 & 1 & $15\{1000\}-425\{820\}+576\{640\}+7568\{622\}-14454\{442\}$  \\
1 & 2 & $35\{911\}-476\{731\}+477\{551\}+204\{533\}$ \\
4 & 0 & $12\{1200\}-456\{1020\}+657\{840\}+13581\{822\}+1137\{660\}-88842\{642\}+114007\{444\}$  \\
0 & 3 & $813\{1200\}{-}30894\{1020\}{+}165652\{840\}{+}72131\{822\}{-}147943\{660\}{-}169702\{642\}{+}57527\{444\}$ \\
\hline
\end{tabular}
}
\end{center}
We note that these are eigenfunctions of the free model, $H_\Om=\half L^2$, since the potential is absent at $g{=}0$ or~$1$,
but they are $S_4$ invariant, The interacting eigenfunctions are of the same form, only the coefficients depend on~$g$.

\section{Tetrahexahedric model: intertwiner and integrability}

In order to construct the intertwiners of the tetrahexahedric model,
one starts with the angular Dunkl operators,
\beq 
\begin{aligned}
\cL_x &\= L_x\ +\ \gg\bigl\{
\sfrac{z}{x-y}s_{x-y}-\sfrac{z}{x+y}s_{x+y}-\sfrac{y}{x-z}s_{z-x}+
\sfrac{y}{z+x}s_{z+x}-\sfrac{y+z}{y-z}s_{y-z}+\sfrac{y-z}{y+z}s_{y+z} \bigr\} \ ,\\
\cL_y &\= L_y\ +\ \gg\bigl\{
\sfrac{x}{y-z}s_{y-z}-\sfrac{x}{y+z}s_{y+z}-\sfrac{z}{y-x}s_{x-y}+
\sfrac{z}{y+x}s_{x+y}-\sfrac{z+x}{z-x}s_{z-x}+\sfrac{z-x}{z+x}s_{z+x} \bigr\} \ ,\\
\cL_z &\= L_z\ +\ \gg\bigl\{
\sfrac{y}{z-x}s_{z-x}-\sfrac{y}{z+x}s_{z+x}-\sfrac{x}{z-y}s_{y-z}+
\sfrac{x}{z+y}s_{y+z}-\sfrac{x+y}{x-y}s_{x-y}+\sfrac{x-y}{x+y}s_{x+y} \bigr\} \ .
\end{aligned}
\eeq
It turns out that the simplest Weyl antiinvariant is cubic,
\beq 
\cM_3 \ \sim\ \sfrac16\bigl(
\cL_x\cL_y\cL_z + \cL_x\cL_z\cL_y + \cL_y\cL_z\cL_x +
\cL_y\cL_x\cL_z + \cL_z\cL_x\cL_y + \cL_z\cL_y\cL_x \bigr)\ ,
\eeq
and taking the Weyl-symmetric reduction we obtain a first angular intertwiner,
\beq 
\begin{aligned}
M_3 \ &\sim\ y^2z\pa_{zxx}-yz^2\pa_{xxy}+\half(y^2{-}z^2)\pa_{xx}+\
4\gg\,\sfrac{yz}{y^2{-}z^2}\bigl(yz\pa_{xx}+x^2\pa_{yz}-zx\pa_{xy})  \\
&+\gg \left[ 2 \gg\,y^2 z^2 \bigl(\sfrac{8 \gg}{\left(x^2-y^2\right) \left(z^2-x^2\right)}+
\sfrac{16 \gg}{\left(z^2-x^2\right) \left(y^2-z^2\right)}-\sfrac{2 \gg-1}{\left(x^2-y^2\right)^2}+
\sfrac{2 \gg-1}{\left(z^2-x^2\right)^2}\bigr)\right.  \\
&\left. \qquad -\sfrac{2 x^2 y^2}{\left(z^2-x^2\right)^2}+\sfrac{2 x^2 z^2}{\left(x^2-y^2\right)^2}-
\sfrac{2 y^2}{x^2-y^2}-\sfrac{2 z^2}{z^2-x^2}-2\sfrac{ y^2+z^2}{y^2-z^2}\right] x \pa_{x} \\
&+2 {\cg g(g{-}1)(g{+}2)}\,x^2 \left[ \sfrac{ y^2+z^2}{\left(y^2-z^2\right)^2}+ z \bigl(\sfrac{1}{(y-z)^3}-
\sfrac{1}{(y+z)^3}\bigr)  \right] +{\cg g\left(2 g^2{+}8 g{-}1\right)}\sfrac{ y^2+z^2}{y^2-z^2} \\
&+2{\cg g^2(8{+}9g)}\sfrac{ x^2 y^2 z^2}{(x^2 - y^2) (x^2 - z^2) (y^2 - z^2)}
-\sfrac{2}{3}{\cg g^3} \sfrac{ x^6+y^6+z^6}{\left(x^2-y^2\right) \left(x^2-z^2\right) \left(y^2-z^2\right)}
+\ \textrm{cyclic permutations}\ .
\end{aligned}
\eeq
In the `potential-free frame', attained by a similarity transformation, it simplifies to
\beq
\begin{aligned}
&\Delta^{-\gg}M_3\,\Delta^\gg\ \sim\ y^2z\pa_{zxx}-yz^2\pa_{xxy}+\half(y^2{-}z^2)\,\pa_{xx}+\
2\gg\,\sfrac{y^2z^2(y^2-z^2)}{(x^2-y^2)(x^2-z^2)}\,\pa_{xx} \\
&+4\gg\,\sfrac{x y^2 z}{x^2-z^2}\,\pa_{xz}
+2 \gg\,x \left[\sfrac{y^2(x^2+3z^2)}{(x^2-z^2)^2} -\sfrac{z^2(x^2+3y^2)}{(x^2-y^2)^2}\right] \partial_{x}
\ +\ \text{cyclic permutations}\ .
\end{aligned}
\eeq
The next independent antiinvariant is sextic,
\beq
\cM_6 \ \sim\
\{\cL_x^4,\cL_y^2\}-\{\cL_y^4,\cL_x^2\}+\{\cL_y^4,\cL_z^2\}
-\{\cL_z^4,\cL_y^2\}+\{\cL_z^4,\cL_x^2\}-\{\cL_x^4,\cL_z^2\}\ ,
\eeq
and gives rise to a rather lengthy expression (not displayed) for a second intertwiner~$M_6$.
We expect that $\Delta^{-\gg}M_6\,\Delta^\gg$ is more compact.
All higher angular intertwiners can be reduced to $M_3$ and~$M_6$.

Let us finally take a look at the conserved charges in this model.
It is not hard to see that they are generated by 
\beq 
J_k \ :=\ \res\bigl(\cL_x^k + \cL_y^k + \cL_z^k\bigr)
\quad\for k=(0,)2,4,6\ ,
\eeq
\beq
\textrm{with}\qquad
J_0 \= C_0 \= 1 \und
J_2 \= -C_2 \= -2\,H_\Om\ +\ 6\gg(6\gg{+}1)\ .
\eeq
Higher conserved charges are algebraically dependent, e.g.
\beq 
\begin{aligned}
6J_8 \ &=\ 8J_6J_2+3J_4J_4-6J_4J_2J_2+J_2J_2J_2J_2 \\
&- 12(8{+}5\gg{+}12{\cg g^2}) J_6 + 4(34{+}23\gg{+}30{\cg g^2}) J_4J_2 - 8(5{+}3\gg{+}3{\cg g^2}) J_2J_2J_2\\
&+ 24(13{+}15\gg{-}102{\cg g^2}{-}72{\cg g^3})J_4 - 4(43{+}70\gg{-}252{\cg g^2}{-}144{\cg g^3}) J_2J_2\\
&- 48(1{+}3\gg)(1{+}4\gg)(1{-}12\gg) J_2\ .
\end{aligned}
\eeq
Any word in $\{J_2,J_4,J_6\}$ is conserved, but there are some relations in their algebra.
Namely, $J_0$ and $J_2$ span the center, and 
\beq
[J_2,J_4] \= [J_2,J_6] \= 0 \qquad\textrm{\cw but}\qquad [J_4,J_6]\ \neq\ 0\ ,
\eeq
so $J_4J_6$ and $J_6J_4$ are two independent new words.
The basic intertwining relations read
{\small
\bea
M_3^{(\gg)} J_2^{(\gg)} \!\!&=&\!\! \bigl(J_2^{({\cg g+1})}-6(7{+}12\gg)\bigr)\,M_3^{(\gg)} \ , \notag\\[4pt]
M_3^{(\gg)} J_4^{(\gg)} \!\!&=&\!\!
\bigl(J_4^{({\cg g+1})}-4(11{+}12\gg)J_2^{({\cg g+1})}+48(26{+}73\gg{+}48{\cg g^2})\bigr)\,M_3^{(\gg)} 
\ +\ 2\,M_6^{(\gg)} \ ,\\[4pt]
M_3^{(\gg)} J_6^{(\gg)} \!\!&=&\!\! \bigl(J_6^{({\cg g+1})}-(35{+}36\gg)J_4^{({\cg g+1})}
-3(7{+}4\gg)J_2^{({\cg g+1})}J_2^{({\cg g+1})} + 2(1111{+}2668\gg{+}1392{\cg g^2})J_2^{({\cg g+1})} \notag \\
&&\!\!+\ 96(457{+}1933\gg{+}2717{\cg g^2}{+}1368{\cg g^3}{+}144{\cg g^4})\bigr)M_3^{(\gg)} 
\ +\ \bigl(3J_2^{({\cg g+1})}-(115{+}200\gg{+}48{\cg g^2})\bigr)M_6^{(\gg)}\ . \notag
\eea
}
Particular conserved quantities are obtained by intertwining `back and forth', e.g.
\beq 
\begin{aligned}
M_3^\+ M_3^{\vphantom{\+}} \ &=\
12J_6-18J_4J_2+6J_2J_2J_2-6(11{+}16 \gg{-}48{\cg g^2})J_4 \\
&+\,3(13{+}24\gg{-}48{\cg g^2})J_2J_2+12(1{+}3\gg)(1{+}4\gg)(1{-}12\gg)J_2 \ ,\\[4pt]
M_6^\+ M_6^{\vphantom{\+}} \ &=\
-12 J_6J_6+12 \{ J_6, J_4\} J_2-\sfrac{16}{3}J_6J_2J_2J_2 +2 J_4J_4J_4-14 J_4J_4J_2J_2 \\
&+\,6J_4J_2J_2J_2J_2-\sfrac{2}{3}J_2J_2J_2J_2J_2J_2 + \ \textrm{lower-order terms}\ ,
\end{aligned}
\eeq
and similarly for \ $M_3^\+ M_6^{\vphantom{\+}}$ \ and \ $M_6^\+ M_3^{\vphantom{\+}}$.
An additional set of `odd' conserved charges appears due to the equality 
$H_\Om^{(g)} = H_\Om^{(1-g)}$ (here $*=3\ \textrm{or}\ 6$):
\beq
Q_{**\cdots*}^{(\gg)} \ := \ M_*^{({\cg g-1})}M_*^{({\cg g-2})}\cdots M_*^{({\cg 1-g})} 
\qquad\cw\Longrightarrow\cy\qquad
Q_{**\cdots*}^{(\gg)} H_\Om^{(\gg)} \= 
Q_{**\cdots*}^{(\gg)} H_\Om^{({\cg 1-g})} \= H_\Om^{(\gg)} Q_{**\cdots*}^{(\gg)}\ .
\eeq
Combining all charges one ends up with a $\Z_2$ graded nonlinear algebra generated 
by \ $\cy \{Q, J_2, J_4, J_6\}$.

\section{Summary and outlook}

Let us summarize. We have presented a geometrical picture of a superintegrable but not
separable potential on $S^{n-2}$. The full set of conserved charges is characterized by
the Weyl invariants built from the Dunkl-deformed angular momenta. Their algebra is
largely unexplored, and it remains to be seen whether there exist bone fide Liouville
charges (i.e.\ $n{-}2$ charges in involution). This angular Calogero system features a 
whole set of angular intertwiners (which also intertwine the full Hamiltonian), given 
by the Weyl antiinvariants built from the angular Dunkl operators. Their form and action 
on the conserved charges was elucidated in the $n{=}3$ (P\"oschl-Teller or hexagonal) 
and $n{=}4$ (tetrahexahedric) cases. For integer coupling there exist additional `odd' 
conserved charges which, however, have a singular action on the energy eigenstates. 
This can be cured by a $\cal PT$ deformation, which regularizes the potential to singular loci 
of codimension two and brings the (so far singular) negative-coupling states into the picture.

\vspace{1cm}

\noindent
{\bf\large Acknowledgments}\\[4pt]
This work was partially supported by
the Alexander von Humboldt Foundation under grant CHL 1153844~STP and
by the Deutsche Forschungsgemeinschaft under grant LE 838/12-2.
This article is based upon work from COST Action MP1405 QSPACE,
supported by COST (European Cooperation in Science and Technology).


\newpage

\begin{thebibliography}{99}

\bibitem{cal71}
F. Calogero,\\
{\it  Solution of a three-body problem in one dimension},\\
J. Math. Phys. {\bf 10} (1969) 2191--2196.

\bibitem{olpe81}
M.A. Olshanetsky, A.M. Perelomov,\\
{\it Classical integrable finite-dimensional systems related to Lie algebras},\\
Phys. Rept. {\bf 71} (1981) 313--400.

\bibitem{olpe83}
M.A. Olshanetsky, A.M. Perelomov,\\
{\it Quantum integrable systems related to Lie algebras},\\
Phys. Rept. {\bf 94} (1983) 313--404.

\bibitem{woj83}
S. Wojciechowski,\\
{\it Superintegrability of the Calogero--Moser system},\\
Phys. Lett. {\bf 95A} (1983) 279--281.

\bibitem{vech90}
O.A. Chalykh, A.P. Veselov,\\
{\it Commutative rings of partial differential operators and Lie algebras},\\
Commun. Math. Phys. {\bf 126} (1990) 597--611.

\bibitem{dun89}
C.F. Dunkl,\\
{\it Differential-difference operators associated to reflection groups},\\
Trans. Amer. Math. Soc. {\bf 311} (1989) 167--183.

\bibitem{hec91}
G.J. Heckman,\\
{\it A remark on the Dunkl differential-difference operators},\\
in: W. Barker, P. Sally (eds.),
{\em Harmonic analysis on reductive groups},\\
Progr. Math. {\bf 101}, 181--191, Birkh\"auser, 1991.

\bibitem{fei03}
M.V. Feigin, \\
{\it Intertwining relations for the spherical parts of generalized Calogero operators},\\
Theor. Math. Phys. {\bf 135} (2003) 497--509.

\bibitem{haneye08}
T. Hakobyan, A. Nersessian, V. Yeghikyan,\\
{\it The cuboctahedric Higgs oscillator from the rational Calogero model},\\
J. Phys. A: Math. Theor. {\bf 42} (2009) 205206 {\tt [arXiv:0808.0430[hep-th]]}.

\bibitem{felepo13}
M. Feigin, O. Lechtenfeld, A. Polychronakos,\\
{\it The quantum angular Calogero-Moser model},\\
JHEP {\bf 1307} (2013) 162 {\tt [arXiv:1305.5841[math-ph]]}.

\bibitem{feha14}
M. Feigin, T. Hakobyan,\\
{\it On Dunkl angular momentum algebra},\\
JHEP {\bf 1511} (2015) 107 {\tt [arXiv:1409.2480[math-ph]]}.

\bibitem{cole15}
F. Correa, O. Lechtenfeld,\\
{\it The tetrahexahedric angular Calogero model},\\
JHEP {\bf 1510} (2015) 191 {\tt [arXiv:1508.04925[hep-th]]}.



\end{thebibliography}
\end{document}